\documentclass[aps,prl,twocolumn, superscriptaddress,showpacs]{revtex4-1}

\usepackage{graphicx}
\usepackage{dcolumn}
\usepackage{bm}
\usepackage{epstopdf}

\usepackage[hyperfigures=true,bookmarksnumbered=true,bookmarksopen=true,bookmarks=true,linkcolor=blue,urlcolor=blue,citecolor=blue,colorlinks=true,pdfhighlight=/O,pdfstartview={XYZ null null 1}]{hyperref}

\begin{document}

\title{Low-temperature lattice effects in the spin-liquid candidate $\kappa$-(BEDT-TTF)$_2$Cu$_2$(CN)$_3$}

\author{Rudra Sekhar Manna}
\email[]{rudra.manna@iittp.ac.in}
\affiliation{Department of Physics, IIT Tirupati, Tirupati 517506, India}
\affiliation{Physics Institute, Goethe University Frankfurt (M), SFB/TR49, D-60438 Frankfurt(M), Germany}

\author{Steffi Hartmann}
\affiliation{Physics Institute, Goethe University Frankfurt (M), SFB/TR49, D-60438 Frankfurt(M), Germany}

\author{Elena Gati}
\affiliation{Physics Institute, Goethe University Frankfurt (M), SFB/TR49, D-60438 Frankfurt(M), Germany}

\author{John A. Schlueter}
\affiliation{Division of Materials Research, National Science Foundation, Arlington, VA 22230, USA}
\affiliation{Materials Science Division, Argonne National Laboratory, Argonne, IL 60439, USA}

\author{Mariano de Souza}
\affiliation{Physics Institute, Goethe University Frankfurt(M), SFB/TR49, D-60438 Frankfurt (M), Germany}
\affiliation{IGCE, Universidade Estadual Paulista, Departamento de Física, Rio Claro, Brazil}

\author{Michael Lang}
\affiliation{Physics Institute, Goethe University Frankfurt(M), SFB/TR49, D-60438 Frankfurt (M), Germany}

\date{\today}


\begin{abstract}

The quasi-two-dimensional organic charge-transfer salt $\kappa$-(BEDT-TTF)$_2$Cu$_2$(CN)$_3$ is one of the prime candidates for a quantum spin-liquid due the strong spin frustration of its anisotropic triangular lattice in combination with its proximity to the Mott transition. Despite intensive investigations of the material's low-temperature properties, several important questions remain to be answered. Particularly puzzling are the 6\,K anomaly and the enigmatic effects observed in magnetic fields. Here we report on low-temperature measurements of lattice effects which were shown to be particularly strongly pronounced in this material (R. S. Manna \emph{et al.}, Phys. Rev. Lett. \textbf{104}, 016403 (2010)). A special focus of our study lies on sample-to-sample variations of these effects and their implications on the interpretation of experimental data. By investigating overall nine single crystals from two different batches, we can state that there are considerable differences in the size of the second-order phase transition anomaly around 6\,K, varying within a factor of 3. In addition, we find field-induced anomalies giving rise to pronounced features in the sample length for two out of these nine crystals for temperatures $T <$ 9 K. We tentatively assign the latter effects to $B$-induced magnetic clusters suspected to nucleate around crystal imperfections. These $B$-induced effects are absent for the crystals where the 6\,K anomaly is most strongly pronounced. The large lattice effects observed at 6\,K are consistent with proposed pairing instabilities of fermionic excitations breaking the lattice symmetry. The strong sample-to-sample variation in the size of the phase transition anomaly suggests that the conversion of the fermions to bosons at the instability is only partial and to some extent influenced by not yet identified sample-specific parameters. 

\end{abstract}

\maketitle

\section{Introduction}

Frustrated magnetism in triangular lattices is one of the growing research interests in condensed matter physics. One class of materials where this physics can be studied are the quasi-two-dimensional organic charge-transfer salts \cite{Toyota2007}. These materials are weak Mott insulators \cite{Toyota2007, Gati2016}, which can be easily converted into a metal, or even a superconductor upon the application of moderate pressure. One of the prime examples is $\kappa$-(BEDT-TTF)$_2$Cu$_2$(CN)$_3$ where the effect of frustration is very strong \cite{Jeschke2012}. This material does not show any long-range magnetic order down to $T$ = 32\,mK \cite{Shimizu2003}, which is four orders of magnitude lower than the estimated nearest-neighbor Heisenberg exchange coupling $J/k_B$ = 250\,K, and has been proposed to be a good candidate for a quantum spin-liquid (QSL) ground state. Although this material has been studied extensively in recent years, there are still several open questions to be answered. A controversial discussion surrounds the nature of the low-lying spin excitations, particularly with regard to the question whether there is a spin gap \cite{MYamashita2009} or not \cite{SYamashita2008}. Another very puzzling issue relates to the so-called 6\,K anomaly. This feature manifests itself in anomalous behavior in various quantities, including $^{13}$C NMR \cite{Shimizu2006}, magnetic susceptibility \cite{Manna2010}, specific heat \cite{SYamashita2008, Manna2010}, thermal conductivity \cite{MYamashita2009}, ultrasound propagation \cite{Poirier2014} as well as thermal expansion \cite{Manna2010}. From the latter experiments, where the strongest response was found, it was claimed that the 6\,K anomaly marks a second-order phase transition. Therefore it may reflect a QSL instability for which various scenarios have been suggested. The proposed models include spin-chirality ordering \cite{Baskaran1989}, a $Z_2$ vortex formation \cite{Kawamura1984}, a pairing of spinons \cite{Lee2007, Galitski2007, Grover2010}, or an exciton condensate \cite{Qi2008}.

Likewise, the influence of a magnetic field on the low-temperature properties of this material confronts us with open questions. On the one hand this relates to the anomalous field-dependent spectral broadening observed in $^{13}$C NMR measurements, which indicates a spatially non-uniform magnetization in this material \cite{Shimizu2006}. On the other hand, the enhancement of the thermal conductivity by the application of magnetic field above 4\,T, was assigned to the $B$-induced closure of a small gap in the magnetic excitation spectrum \cite{MYamashita2009, MYamashita2012}. Moreover, the existence of a $B$-induced quantum phase transition at a very small field of about 5\,mT was claimed from results of $\mu$SR experiments \cite{Pratt2011}. It was argued that this quantum phase transition separates a gapped spin liquid phase, with a tiny spin gap of $\Delta_s/k_B \sim$ 3.5\,mK, from a weak-moment antiferromagnetic phase. According to these studies, a second quantum critical point exists in this material around 4\,T which was assigned to a threshold for deconfinement of spin excitations \cite{Pratt2011}. 

In light of these intriguing field-dependent effects and the complex phenomenology in zero field, one may ask about sample-to-sample variations of the material's properties. In fact, indications for considerable sample dependences were found in thermal conductivity measurements \cite{MYamashita2009}. Here, we report an extensive study of sample-to-sample variations of the low-temperature behavior by focussing on the lattice effects around the 6\,K phase transition. We find large variations of the size of the phase transition anomaly in the coefficient of thermal expansion, up to a factor of 3, whereas its position varies only slightly around 6\,K. In addition, for two crystals out of nine, we find highly anomalous lattice effects when a magnetic field is applied along the in-plane $b$-axis.

\section{Experimental}

Single crystals with typical dimensions of about 0.1$\times$1.0$\times$1.2 mm$^3$ were used for the experiments. The crystals were grown by following the standard procedure described in Ref.\,\cite{Geiser1991}. An ultrahigh-resolution capacitive dilatometer was employed for the thermal expansion measurements (built after \cite{Pott1983}), enabling the detection of length changes $\Delta l \geq$ 10$^{-2}$ \AA, where $l$ is the length of the sample. For measurements in a constant magnetic field as a function of temperature and also for measurements of the magnetostriction at constant temperature, the magnetic field was applied along the measuring direction of the crystal. Thermal expansion measurements at 0 $\leq B \leq$ 10\,T were performed upon heating and cooling with a slow sweep rate of $\pm$1.5 K/h to ensure thermal equilibrium. For magnetostriction measurements, the sweep rate of the magnetic field was $\pm$120 mT/min. Overall nine single crystals were studied, five single crystals from batch no.\,KAF 5078 and four from batch no.\,MP 1049.

\section{Sample-to-sample variations of the 6\,K anomaly}

\begin{figure}[!htb]
\centering
\includegraphics[width=0.80\columnwidth]{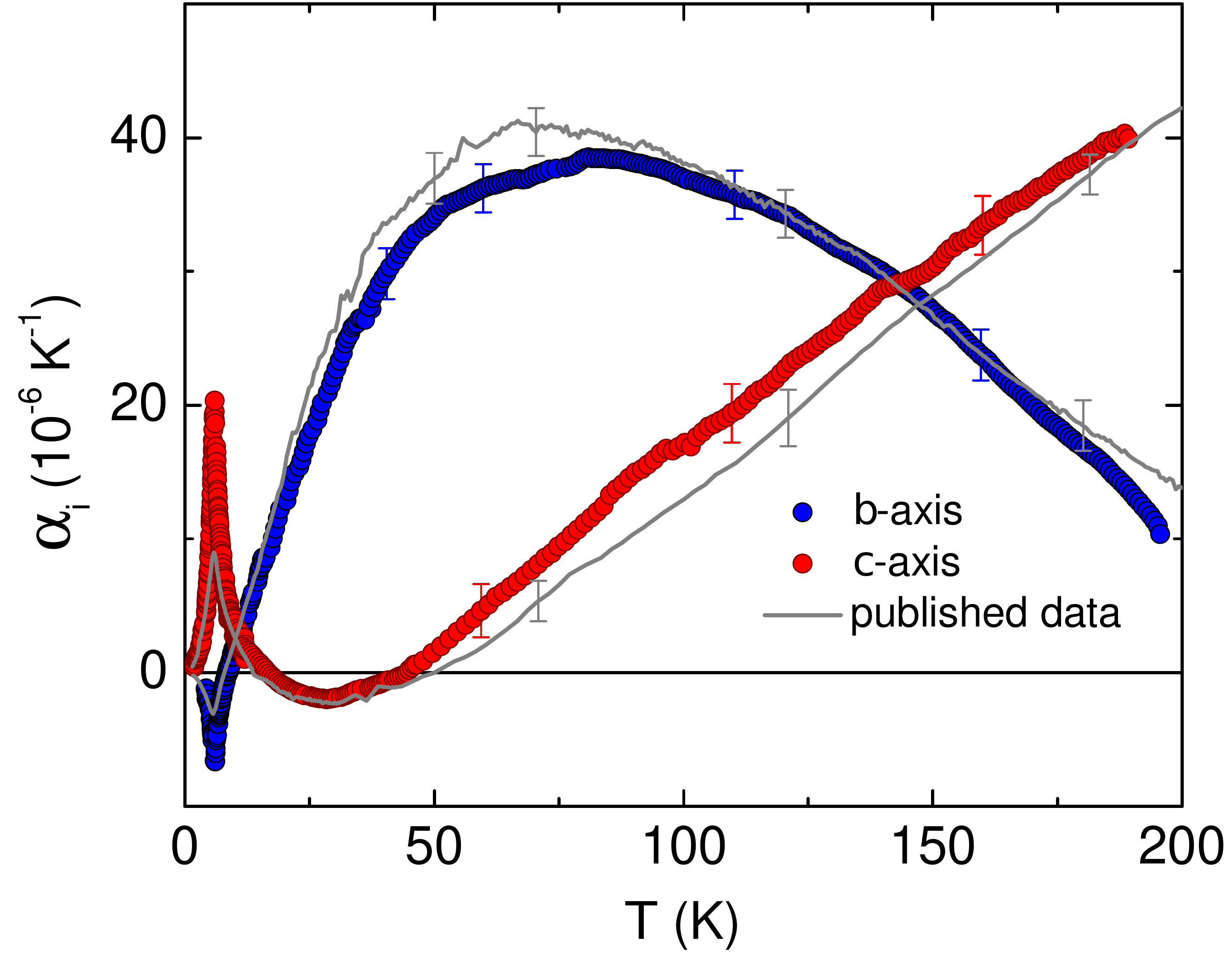}
\caption{\label{overview} Overview of the coefficients of thermal expansion for $\kappa$-(BEDT-TTF)$_2$Cu$_2$(CN)$_3$ single crystal MP 1049\#2 (symbols) measured along the in-plane $b$- and $c$-axis for $T \leq$ 200\,K. The solid gray line corresponds to data for single crystals from batch KAF 5078 reported previously by Manna \emph{et al.} \cite{Manna2010}.}
\end{figure}

Figure \ref{overview} gives an overview of the in-plane thermal expansion coefficients $\alpha_i$($T$) = $l_i^{-1} \partial l_i$($T$)/$\partial T$ ($i$ = $ b$, $c$ are the in-plane crystallographic axes) for the $\kappa$-(BEDT-TTF)$_2$Cu$_2$(CN)$_3$ single crystal MP 1049\#2 (symbols) over the whole temperature range investigated. For comparison, we show in Fig. \ref{overview} corresponding data for the crystals discussed in Ref.\,\cite{Manna2010} (gray line) which were taken from batch KAF 5078. Besides the sharp peaks in $\alpha_b$ and $\alpha_c$ around 6\,K, which will be discussed below in more detail, the data at higher temperatures reveal highly anomalous and strongly anisotropic behavior. This includes a pronounced maximum of $\alpha_b$ around 80\,K which is absent for $\alpha_c$. Instead, $\alpha_c$ decreases almost linearly with decreasing temperatures, becomes negative below about 50\,K and passes through a minimum around 30\,K. We stress that, apart form the 6\,K anomaly, the $\alpha_b$ and $\alpha_c$ data are almost identical, within the experimental resolution, to those revealed for crystals from batch KAF 5078 \cite{Manna2010} (solid gray line in Fig.\,\ref{overview}). As was discussed in Ref.\,\cite{Manna2010}, the anomalies in $\alpha_b$ and $\alpha_c$ indicate that besides phonons, also other excitations contribute substantially to the low-temperature thermal expansion of this material.

\begin{figure}[!htb]
\centering
\includegraphics[width=0.8\columnwidth]{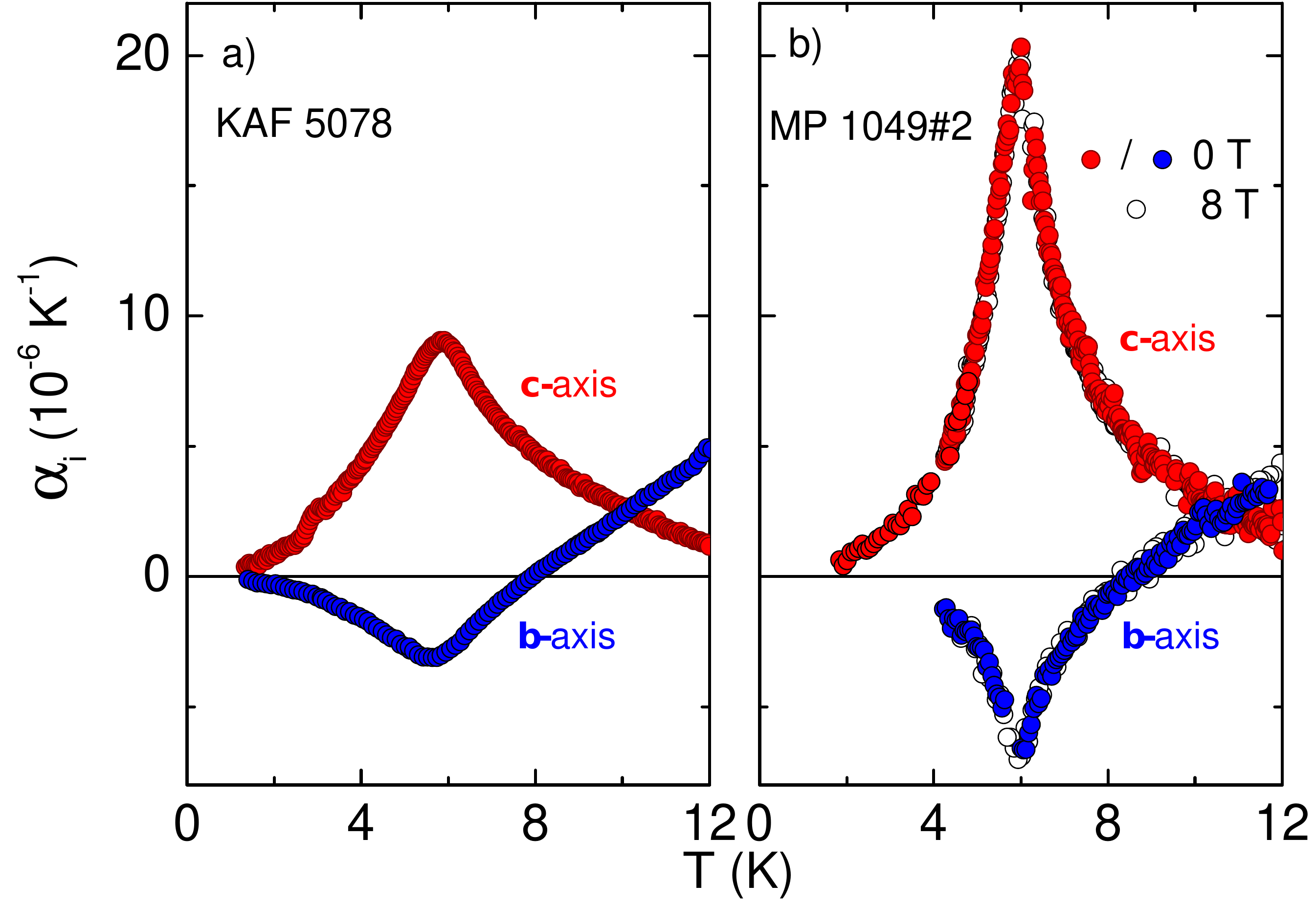}
\caption{\label{6K-bc-comparison} Thermal expansion coefficients for $\kappa$-(BEDT-TTF)$_2$Cu$_2$(CN)$_3$ measured along the in-plane $b$- and $c$-axis around the 6\,K phase transition: (a) on single crystal from batch KAF 5078 reported in \cite{Manna2010}, (b) on single crystal MP 1049\#2 studied in this work, where the anomalies are most strongly pronounced. For this crystal also data taken in a magnetic field of $B$ = 8\,T applied along the measuring direction are shown.}
\end{figure}

The low-temperature thermal expansion coefficients are dominated by the 6\,K anomaly yielding sharp spikes in $\alpha_b$ and $\alpha_c$ with reversed sign. The data for $T \leq$ 12\,K are shown in Fig.\,\ref{6K-bc-comparison}(b) on enlarged scales. For comparison, we show in Fig.\,\ref{6K-bc-comparison}(a) the corresponding data for the single crystals from batch KAF 5078 reported previously by Manna \emph{et al.} \cite{Manna2010}.

\begin{figure}[!htb]
\centering
\includegraphics[width=0.6\columnwidth]{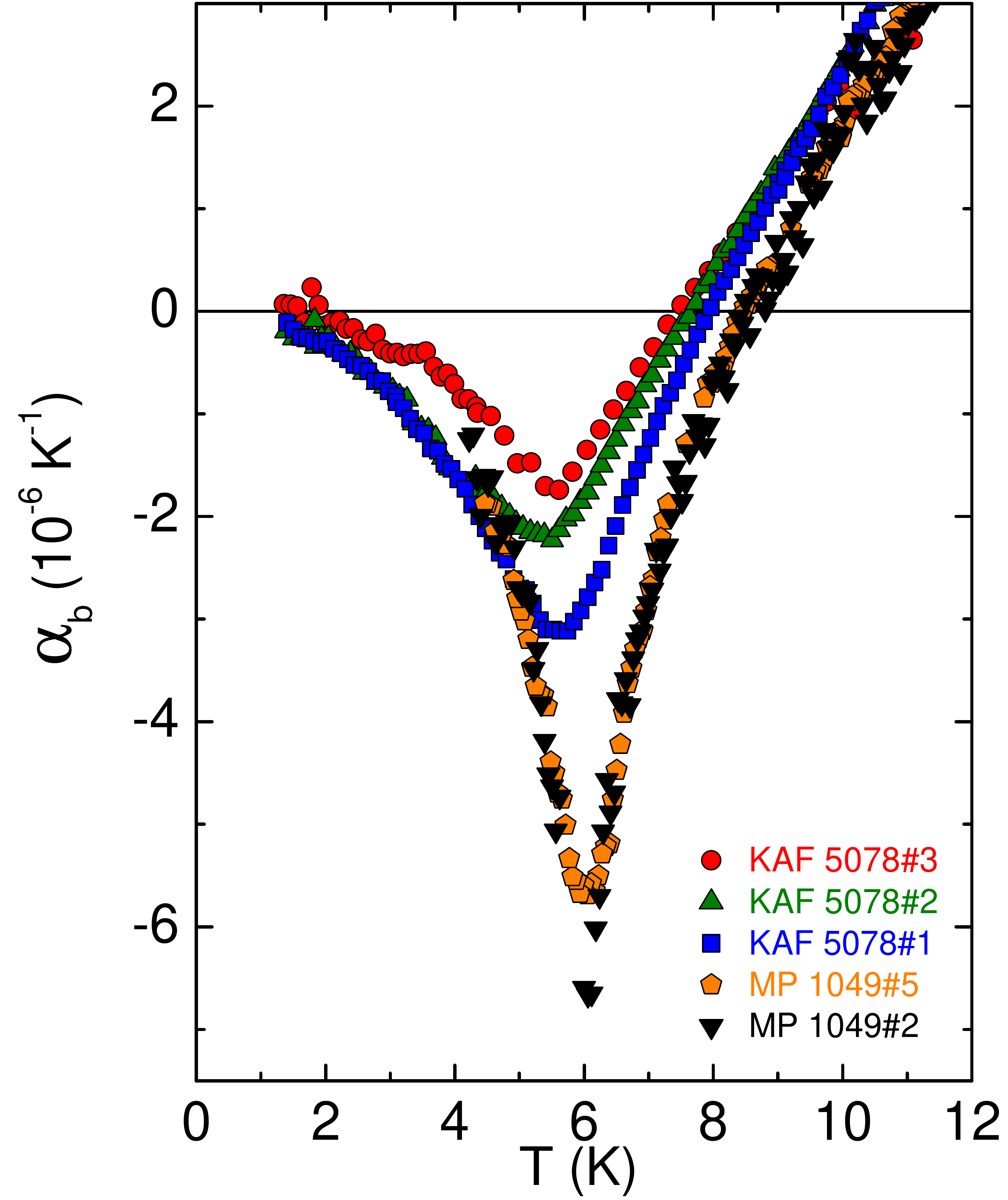}
\caption{\label{6K-b-comparison} Comparison of the in-plane $b$-axis thermal expansion coefficient of $\kappa$-(BEDT-TTF)$_2$Cu$_2$(CN)$_3$ for a selection of 5 out of 9 single crystals taken from two different batches.}
\end{figure}

Figure \ref{6K-bc-comparison} discloses a strongly sample-dependent anomaly at 6\,K. For crystal MP 1049\#2, the size of the peaks in $\alpha_b$ and $\alpha_c$ are not only about two times larger than the ones found earlier on single crystals from batch KAF 5078 \cite{Manna2010}. The anomalies are also distinctly sharper and more asymmetric in temperature with a steeper flank on the low-temperature side of the peak, clearly identifying the feature as a second-order phase transition. Despite these differences, however, other characteristics of the transition are retained. This includes the peak position at $T_p \simeq$ 6\,K, the anisotropy ratio $\alpha_c$($T_p$)/$\alpha_b$($T_p$) $\sim$ 3, and a crossing point of $\alpha_b$ and $\alpha_c$ at around 10\,K. To illustrate the extent this sample-to-sample variation can take, we show in Fig.\,\ref{6K-b-comparison} a compilation of $\alpha_b$ data for five selected single crystals from two different batches, including the crystals KAF 5078\#1 and MP 1049\#2 presented above in Fig.\,\ref{6K-bc-comparison}. Figure \ref{6K-b-comparison} discloses a huge variation by a factor of about 3 in the size of the transition, whereas the position changes only slightly within about 0.5\,K. Note that, even though the largest difference occurs between crystals from the different batches, there are also strong variations for crystals from the same batch. 

\section{Field-induced effects}

\begin{figure}[!htb]
\centering
\includegraphics[width=0.8\columnwidth]{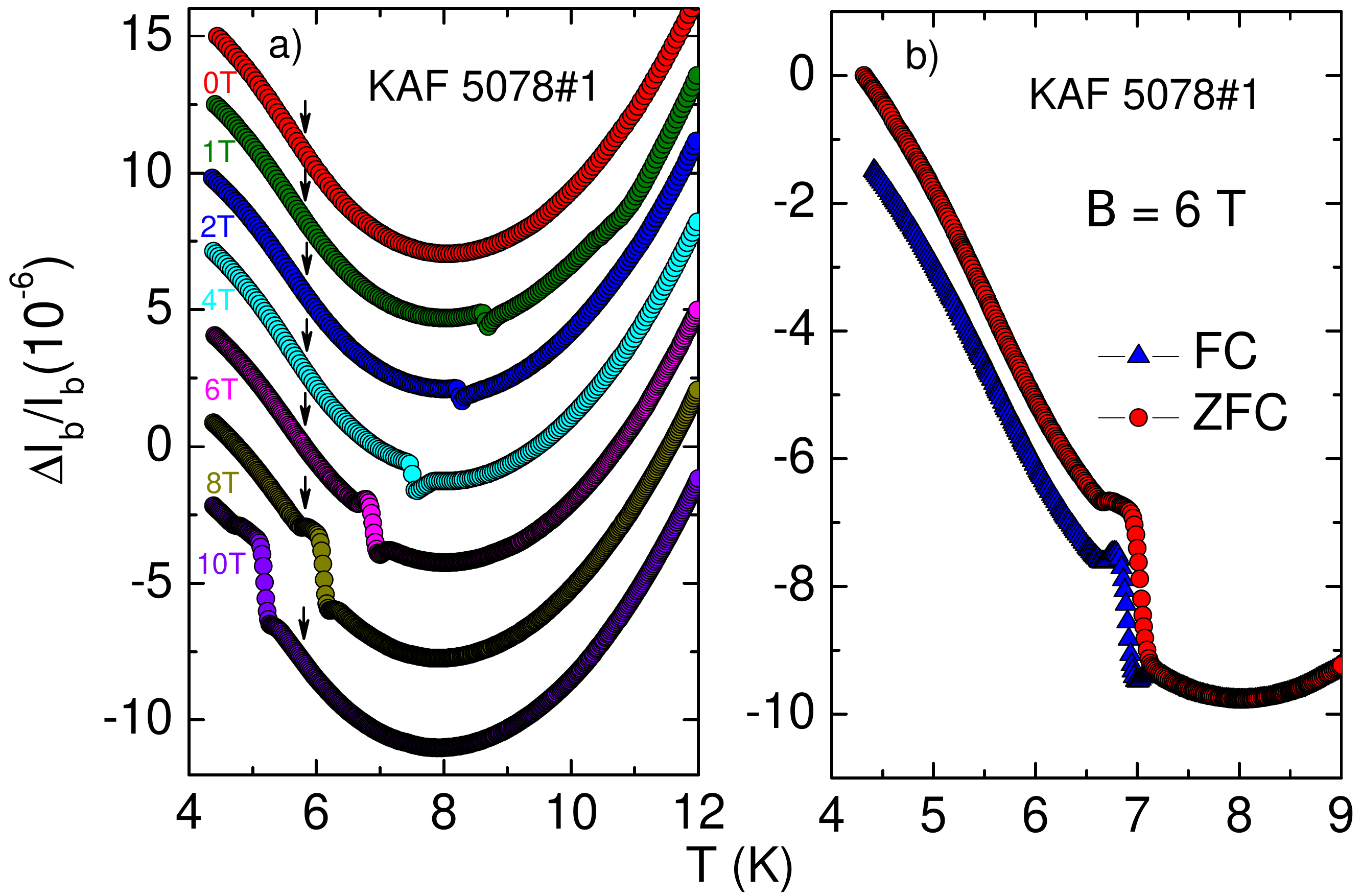}
\caption{\label{field-effect} (a) Temperature-dependent relative length changes, $\Delta l_b/l_b$, for $\kappa$-(BEDT-TTF)$_2$Cu$_2$(CN)$_3$ (KAF 5078\#1) along the $b$-axis for various constant magnetic fields between 0\,T to 10\,T. The curves were shifted along the $y$-axis for clarity. The arrows indicate the phase transition at 6\,K for the various fields, giving rise to a peak in the thermal expansion coefficient, $\alpha_b$. (b) Relative length changes for zero-field cooling (ZFC) and field cooling (FC) at a constant field value of 6\,T. Measurements were performed with a very slow rate of $\pm$1.5\,K/h.}
\end{figure}

All nine crystals, including the ones shown in Fig.\,\ref{6K-b-comparison}, were also subject to measurements in magnetic fields. The following observations were made: 1) as shown in Fig.\,\ref{6K-bc-comparison}, there is no obvious effect of a magnetic field up to 8\,T, the maximum field applied, on the anomaly in $\alpha_c$ ($B \parallel c$-axis). For this crystal MP 1049\#2, this statement is true also for $\alpha_b$ ($B \parallel b$-axis). In these experiments the field was applied at a temperature of 12\,K, prior to the measurements. 2) In contrast, for two crystals (KAF 5078\#1 and KAF 5078\#4) out of all nine crystals studied, we find highly anomalous $B$-induced lattice effects when $B$ is applied along the $b$-axis of the crystal. 
The $B$-induced anomalous behavior is shown in the left panel of Fig.\,\ref{field-effect} where we plot the relative length changes $\Delta l_b(T)/l_b$ = $l(T_0)^{-1}\cdot$[$l(T)-l(T_0)$], with $T_0$ a reference temperature, along the in-plane $b$-axis as a function of temperature at different constant magnetic fields applied parallel to the $b$-axis, see Ref.\,\cite{RManna2012} for a preliminary report of the investigations. For comparison, we include in Fig.\,\ref{field-effect}(a) the data taken at zero magnetic field, yielding a broad minimum at around 8\,K, which corresponds to the change of sign of $\alpha_b$ = $l_b^{-1}\partial l_b/\partial T$ (Fig.\,\ref{6K-bc-comparison}). On the scale of Fig.\,\ref{field-effect}, the abrupt change in slope in the $\Delta l_b/l_b$ data at 6\,K (indicated by an arrow), reflecting the pronounced phase transition anomaly in $\alpha_b$ (Fig.\,\ref{6K-bc-comparison}), cannot be seen. The same results, without any obvious field-induced anomaly, were obtained in a field of $B$ = 0.5\,T \cite{RManna2012} (not shown). However, upon increasing the field to $B$ = 1\,T, the data reveal a jump-like anomaly at 8.7\,K. The anomaly grows in size and shifts to lower temperatures down to 5.2\,K with increasing magnetic fields up to 10\,T, the highest field accessible. These results suggest that a field in excess of some threshold value 0.5\,T < $B$ < 1\,T is necessary to trigger this effect. Interestingly, the magnetic field does not affect the 6\,K phase transition anomaly. These measurements were performed upon cooling with a rate $-$1.5\,K/h and the magnetic field was applied at 12\,K. We stress that measurements along the second in-plane $c$-axis with field parallel to $c$ \cite{Manna2010} and measurements along the out-of-plane $a$-axis with field parallel $a$ \cite{Manna2012} failed to find any indication for such a field-induced anomaly.

Irrespective of the fact that the field-induced anomalies were seen only in two out of nine crystals, it is enlightening to explore the phenomenology of these anomalies in more detail. At first glance, one would be inclined to assign the discontinuous length changes revealed in $B \geq$ 1\,T to a first-order phase transition. However, the absence of any hysteresis in $\Delta l_b/l_b$ upon heating and cooling with a slow rate of $\pm$1.5 K/h \cite{Manna2012} speaks against such an interpretation. Likewise, changing the heating and cooling rates (from $\pm$0.5 K/h to $\pm$5.0 K/h) were found to have no effect on the anomaly (not shown) which is an indication that there is no spin-glass behavior involved. Furthermore, as was shown in Ref.\,\cite{RManna2012}, a comparison of $\Delta l_b/l_b$ data from 4.5\,K to 12\,K, between zero field and a finite field of 6\,T, reveals that the data lie on top of each other at the high- and low-temperature end, but significantly deviate from each other at intermediate temperatures. This suggests that the jump-like anomaly in the intermediate region indicates a release of a field-induced lattice strain upon cooling \cite{RManna2012}. Whereas there is no hysteresis upon heating and cooling, we do find a significant difference in $\Delta l_b/l_b$ between zero-field cooling (ZFC) and field cooling (FC) experiments, cf.\, Fig.\,\ref{field-effect}(b). In the experiments shown there, the sample was zero-field cooled down to 4.5\,K, a field of 6\,T was applied, and then data were taken upon heating (red circles) at a rate of $+$1.5 K/h (ZFC). With a delay of one night, the second data set was taken where the field was applied at 12\,K and data were taken upon slowly cooling (blue triangles) with a rate of $-$1.5 K/h (FC). In the figure, the data sets were shifted vertically so that they coincide at the high-temperature end. 

\begin{figure}[!htb]
\centering
\includegraphics[width=0.8\columnwidth]{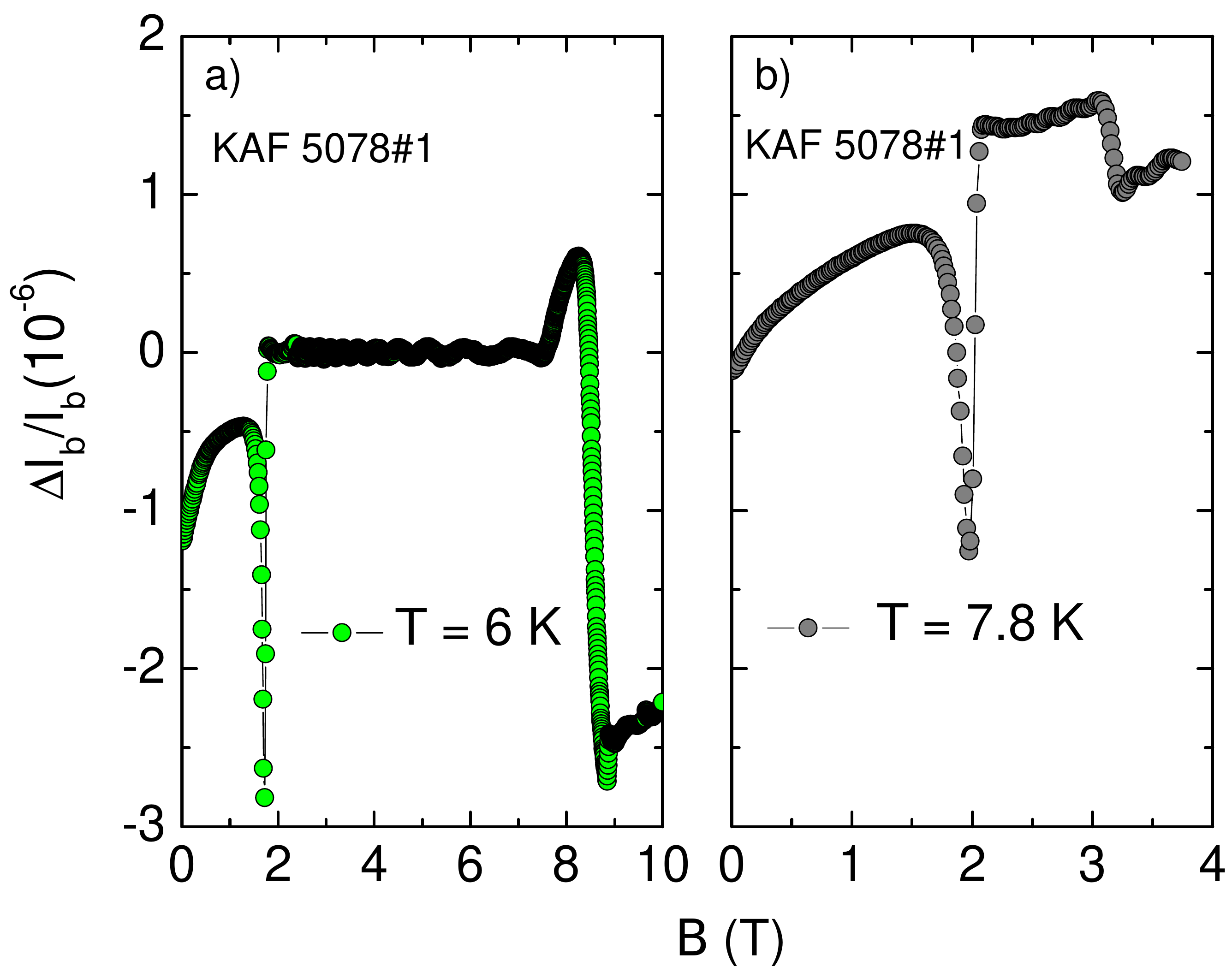}
\caption{\label{magnetostriction} Relative length changes along the $b$-axis, $\Delta l_b/l_b$, for $\kappa$-(BEDT-TTF)$_2$Cu$_2$(CN)$_3$ (KAF 5078\#1) as a function of applied magnetic field, $B \parallel b$, at a temperature (a) $T$ = 6\,K and (b) $T$ = 7.8\,K. The oscillations in the data, being periodic in $B^{-1}$, are due to quantum oscillations of gallium ($Ga$) used for affixing the sample in the desired orientation.} 
\end{figure}

In addition to the temperature-dependent investigations in constant fields, we have looked for corresponding anomalies also in magnetostriction experiments, i.e., measurements of $\Delta l_b/l_b$ upon varying the magnetic field up to 10\,T. The measurements were performed by employing a sweep rate of $\pm$120 mT/min. In the following we discuss a selection of the magnetostriction results. In Fig.\,\ref{magnetostriction}(a), we show the relative length changes along the $b$-axis as a function of magnetic field ($B \parallel b$) at $T$ = 6\,K. The data reveal a pronounced step-like anomaly slightly above 8\,T which corresponds to the feature observed in temperature sweeps at $B$ = constant. Interestingly enough, these magnetostriction measurements reveal yet another anomaly at a lower field around 1.8\,T which could not be seen in temperature sweeps at constant fields. Corresponding data for temperature $T$ = 7.8\,K are shown in Fig.\,\ref{magnetostriction}(b). Similar to the data at 6\,K, we find two anomalies, a sharp peak-like feature, now located around 2\,T, and a step-like feature at higher fields.

\begin{figure}[!htb]
\centering
\includegraphics[width=0.80\columnwidth]{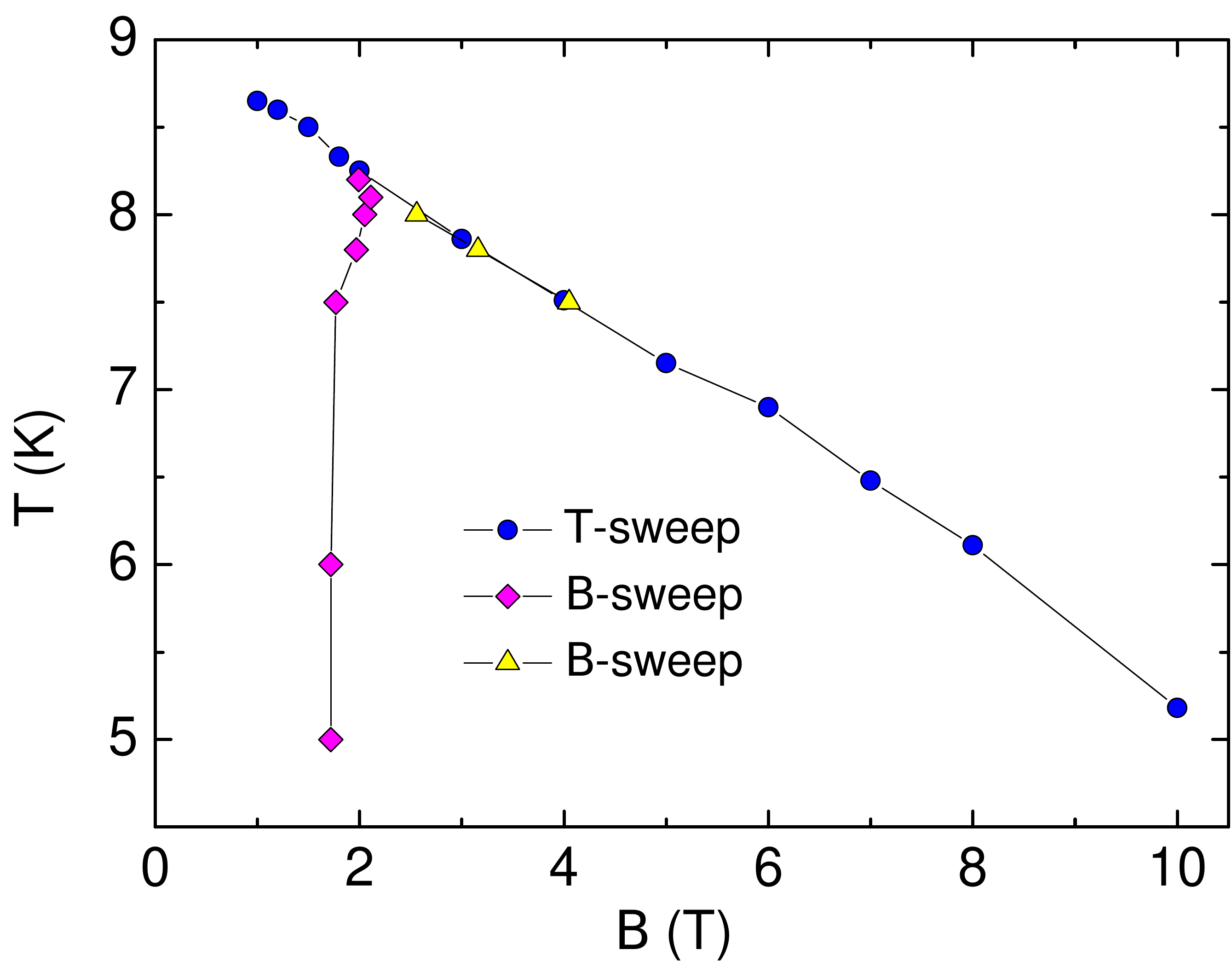}
\caption{\label{anomaly-diagram} Anomaly diagram for $\kappa$-(BEDT-TTF)$_2$Cu$_2$(CN)$_3$ (KAF 5078\#1) based on the position of the field-induced anomalies in $\Delta l_b/l_b$ as determined from thermal expansion measurements at $B$ ($\| b$) = constant ($T$-sweep, blue circles) and magnetostriction at $T$ = constant ($B$-sweep, magenta diamonds and yellow triangles). We note that these effects are absent in measurements along both the in-plane $c$-axis with $B \| c$ and along the out-of-plane $a$-axis with $B \| a$.}
\end{figure}

Based on results from thermal expansion measurements as a function of temperature ($T$-sweep) at constant fields and results from magnetostriction measurements for isothermal field sweeps ($B$-sweep), an anomaly diagram can be constructed, as shown in Fig.\,\ref{anomaly-diagram}. The position of the anomaly at higher fields, derived from magnetostriction measurements, are fully consistent with those revealed from measurements as a function of temperature at $B$ = constant. Two distinct magnetic field-induced features can be identified which are most strongly pronounced and well-separated from each other at low temperatures, while the anomalies merge together at around 8.4\,K. A finite field above some threshold value of 0.5\,T $< B_c\leq$ 1\,T is necessary to observe these anomalous field-dependent effects. 

\section{Discussion}

The phenomenology described above suggests that the field-induced anomalies do not reflect equilibrium properties of the hypothetically ideal material. In an attempt to provide an interpretation of these effects, we recall that (i) there is a significant sample-to-sample variation in the occurrence of the $B$-induced anomalies, and (ii) there is no obvious interrelation with the 6\,K anomaly. The latter statement is based on the following two observations: there is a continuous, smooth growth of the $B$-induced anomaly on increasing the field from 6\,T over 8\,T to 10\,T (Fig.\,\ref{field-effect}), despite crossing the phase boundary associated to the 6\,K anomaly, (cf.\,Fig.\,\ref{anomaly-diagram}). In other words, the two effects interpenetrate each other as a function of field without any mutual influence. In addition, there seems to be an anticorrelation with the 6\,K anomaly: the $B$-induced effects are absent in those crystals where the 6\,K anomaly is strongest pronounced.

We are inclined to assign these effects to a $B$-induced formation of local magnetization which may nucleate around impurities or grain boundaries, as suggested on the basis of NMR measurements \cite{Shimizu2006}. We further suspect that for the two crystals (\#1 and \#4 from batch KAF 5078) around those sites and induced by a finite field, some kind of small antiferromagnetic clusters are formed with an easy axis parallel to the $b$-axis. We then assign the spike-like feature around $B \simeq$ 1.8-2\,T to the spin-flop transition of these antiferromagnetic clusters. In fact, the anomaly diagram presented in Fig.\,\ref{anomaly-diagram} bears some resemblance to that of an uniaxial antiferromagnet with the field applied parallel to the preferred axis of spin alignment. In those uniaxial antiferromagnets, the transition from the antiferromagnetic phase to the spin-flop phase is of first order and almost independent of temperature, while the transition from the spin-flop phase to the fully polarized state at higher fields is of second order, showing a strong temperature dependence. The fact that we observe jump-like features at higher fields, and a ZFC-FC hysteresis (cf.\,Fig.\,\ref{field-effect}(b)), not expected for simple uniaxial antiferromagnets, is presumable due to the small, very likely nano-scale size of the $B$-induced magnetic clusters. As known from studies on magnetic nano-structures, the increased surface contribution of these structures can create irreversible contributions to the magnetization, even though the core of these structures is antiferromagnetic, see, e.g. Ref.\,\cite{Benitez2008}.

The fact that these $B$-induced anomalies are absent for the crystals MP 1049\#2 and MP 1049\#5 is consistent with the above interpretation and supports the view that the crystals, where the 6\,K anomaly is most strongly pronounced, have a lower concentration of those defects on which magnetization can nucleate. Although the present study cannot make a definite statement about the nature of the 6\,K transition, it clearly demonstrates that the order parameter strongly couples to the lattice degrees of freedom. Hence our results are consistent with models \cite{Lee2007, Grover2010, Qi2008} predicting a QSL instability that breaks the lattice symmetry so that pronounced lattice effects are expected. The pairing of fermions (spinons or excitons), considered in these models, giving rise to a conversion to bosons, may be partial due to intrinsic but also extrinsic reasons. The high sensitivity of the size of the 6\,K phase transition anomaly to some (yet unknown) sample-specific parameters, may then correspond to a different fraction of the fermions forming bosonic pairs at $T_p$ = 6\,K. In light of the variation in the size of the anomaly within a factor of about 3, we expect that there could be a considerable sample-to-sample variation in this fraction. Depending on the sample investigated and the experimental probe applied, this may lead to quite different conclusions as for the character of excitations of the low-temperature state, and therefore could provide a plausible explanation for the ongoing controversy on these issues.

\section{Summary}

In summary, detailed investigations of low-temperature lattice effects have been performed on the proposed spin-liquid compound  $\kappa$-(BEDT-TTF)$_2$Cu$_2$(CN)$_3$. Particular emphasis was placed on sample-to-sample variations around the mysterious 6\,K anomaly and the enigmatic field effects. By studying overall nine crystals from two different batches we found that the second-order phase transition at 6\,K is strongly sample dependent in its size, varying within a factor of 3, whereas the position stays constant within 0.5\,K. In two out of these nine crystals, we observe pronounced field-induced effects, which were tentatively assigned to the formation of small antiferromagnetic clusters suspected to nucleate around some crystal imperfections. These effects are absent for those crystals where the phase transition anomaly at 6\,K is most strongly pronounced. Our results are consistent with a pairing instability of the quantum spin liquid at 6\,K which breaks lattice symmetry. We suspect that the conversion of fermionic excitations to bosons at this transition is only partial and to some extent influenced by sample-dependent factors. 

\section{Acknowledgments}

We acknowledge financial support by the Deutsche Forschungsgemeinschaft via the SFB/TR49 and Claudius Gros, Roser Valenti, Patrick A. Lee, Jens M\"{u}ller for useful discussions. RSM acknowledges the financial support from IIT Tirupati. Work at Argonne National Laboratory (ANL) was supported by UChicago Argonne, LLC, Operator of ANL. Argonne, a U.S. Department of Energy Office of Science laboratory, is operated under contract no. DE-AC02-06CH11357. JAS acknowledges support from the Independent Research/ Development program while serving at the National Science Foundation.

\section{author contributions}

Measurements were performed by R.S.M. and S.H. with contributions from M.S. and E.G. Single crystals were grown by J.A.S. R.S.M. and M.L. wrote the paper. All authors discussed the results and commented on the manuscript.

\section{conflicts of interest}

The authors declare no conflict of interest.

\end{document}